\documentclass[11pt,twoside]{article}
\usepackage{cspm-asp2006}
\usepackage{epsfig,graphicx,natbib,url}  
\usepackage{lscape}
\begin{document}

\markboth{Z.~F. Bostanc\i~\& N. Al Erdo\u{g}an}{Temporal evolution of mottles observed in H$\alpha$}
\title{Temporal evolution of mottles observed in H$\alpha$}
\author{Z. Funda Bostanc\i$^{1}$ and  
             Nurol Al Erdo\u{g}an$^1$}
\affil{$^1$Istanbul University, Faculty of Science, Department of Astronomy \& Space Sciences, 34119 University-Istanbul, Turkey}

\begin{abstract}
In April 2002, H$\alpha$ observations of the solar chromosphere with high spatial 
and spectral resolution were obtained with the \textquoteleft G\"ottingen Fabry-Perot Spectrometer\textquoteright~mounted in the Vacuum Tower Telescope (VTT) at the Observatorio del Teide/Tenerife. 
In this work, we analyze a short time sequence of a quiet region with chains of mottles. 
Some physical parameters of dark mottles are determined by using Beckers' cloud model
which takes the source function, the Doppler width, and the velocity to be constant 
within the cloud along the line of sight. Here, we present the results of our study.
\end{abstract}

\section{Introduction}

Two dimensional H$\alpha$ observations of network regions on the solar disk center with high spatial resolution show dark elongated structures, the so-called \textquoteleft dark mottles\textquoteright. These structures usually form two groups, which are called chains of mottles or rosettes (Beckers, 1963). In a chain, mottles point in the same direction while rosettes have a more or less circular shape with a bright center, which is surrounded by a number of mottles aligned radially outwards \citep{tziotziou03}.

Beckers' cloud model \citep{beckers64} is widely used for the determination of physical quantities of mottles such as the line-of-sight velocity, the source function, the optical depth, and Doppler width \citep{tsiropoula94, lee00, tziotziou03, al04, bostanci05}. Here, we applied this model to a short time series of a two-dimensional field of view which includes chains of mottles. 

\section{Observations \& Data Analysis}
The observation of a network region near disk center was
done in H$\alpha$
using the \textquoteleft G\"ottingen Fabry-Perot Spectrometer\textquoteright~mounted in the Vacuum
Tower Telescope (VTT) at the Observatorio del Teide/Tenerife
\citep{bendlin95, koschinsky01}. Two-dimensional broad-band and
narrow-band images were taken simultaneously at 18 wavelength
positions by scanning through H$\alpha$. The wavelength settings of
two consecutive positions differed by 125 m\AA\, and the time interval between two consecutive scans was 49 s. The broad-band images were restored by spectral ratio \citep{luhe84} and speckle masking \citep{weigelt77} methods while the narrow-band images were reconstructed by using a method given by Keller \& von der L\"uhe (1992). The H$\alpha$ line profile for each pixel of the field of view was produced from intensity values of narrow-band images belonging to 18 wavelength positions.

\begin{figure}
  \centering
  \hbox{ \hspace{0.0in}}
  \includegraphics[scale=0.92]{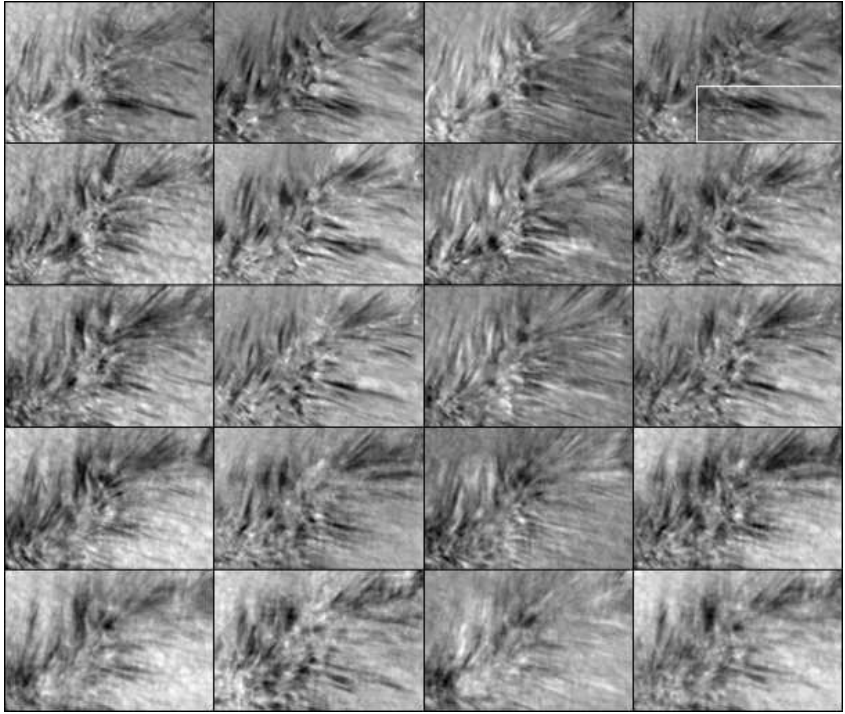}
  \caption[]{\label{fb-fig:timeseries}
A short time sequence of the entire FOV (30\farcs3$\times$20\farcs5) at
 +0.650~\AA\,(first column) and $-$0.650~\AA\,(second column) from the
 H$\alpha$ line center with corresponding Doppler and
 intensity images (third and fourth column, respectively). The time cadence is 49 s.
}\end{figure}

\section{Cloud Model}
The observed contrast profiles were matched
with theoretical contrast profiles by using the cloud model \citep{beckers64}. This model assumes the source function, the Doppler width, and the LOS velocity to be constant within the cloud along the line of sight. The contrast profile is given by
\begin{equation}
C(\lambda)={{I(\lambda)-I_0(\lambda)}\over{I_0(\lambda)}}=\left({S\over{I_0(\lambda)}}-1\right)(1-\rm{e}^{{-}\tau(\lambda)})
\label{eq:equation1}
\end{equation}
where $I_0(\lambda)$ is the reference profile emitted by the background. S denotes the source function and $\tau(\lambda)$ the optical thickness. The wavelength dependence of the optical thickness is given by 
\begin{equation}
\tau(\lambda)=\tau_0~\rm{\exp}\left[-\left({\lambda-\lambda_c(1-\upsilon_{\rm{LOS}}/c)\over \Delta\lambda_D}\right)^2\right]
\label{eq:equation1}
\end{equation}
where $\tau_0$ is the optical depth at the line center, $\lambda_{\rm{c}}(\upsilon_{\rm{LOS}}/c)$ the shift due to the velocity $\upsilon_{\rm{LOS}}$ and $\lambda_{\rm{c}}$ the line center wavelength.

\begin{figure}
  \centering
  \includegraphics{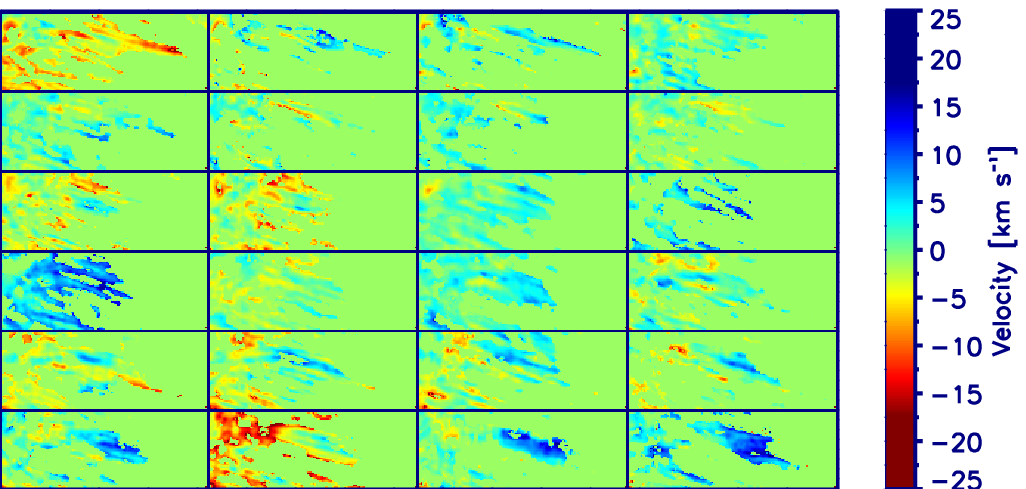}
  \includegraphics{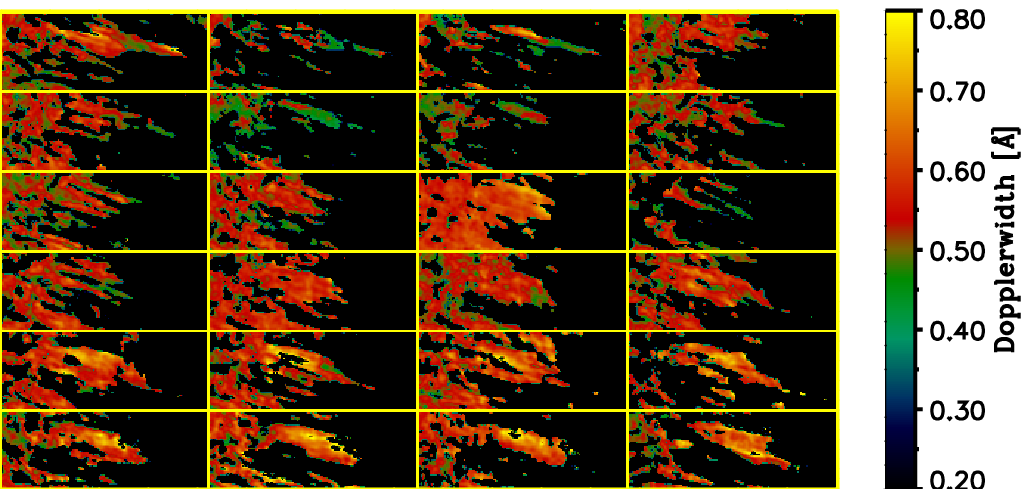}
  \includegraphics{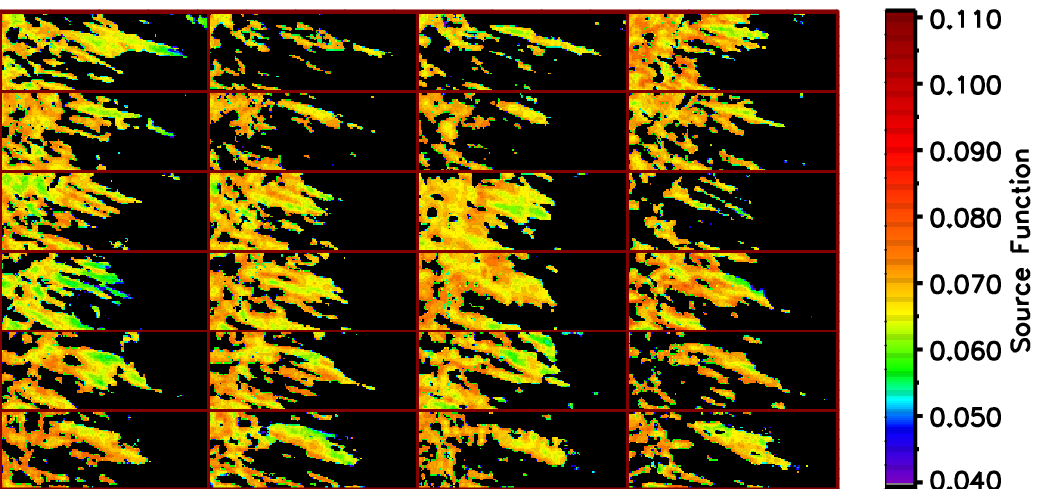}
  \includegraphics{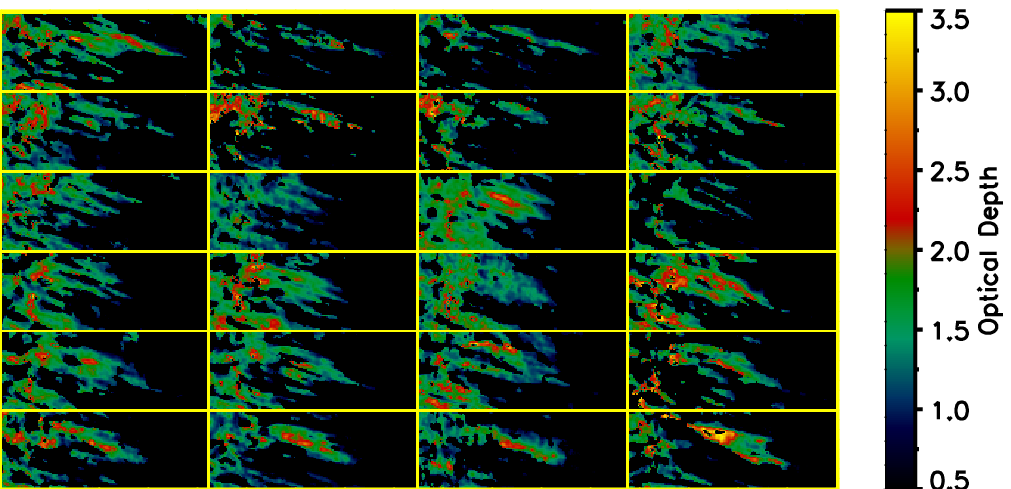}
  \caption[]{\label{fb-fig:parameterimages}
Short time sequences of velocity, Doppler width, source function, and
optical depth for the region marked by the white frame in Figure 1 (corresponding to the size of
 21\farcs3$\times$8\farcs1). Time proceeds row by row from top to bottom with a cadence of 49 s. In the
 velocity images, blue and red colours indicate upward and downward
motion, respectively.
}\end{figure}

\section{Results}
In this study, we found the average thickness of the mottles to be
0\farcs7 ranging between 0\farcs3 and 1\farcs1 and the average length to be 9\farcs8
ranging between 6\farcs7 and 13\farcs4. Their lifetime is close to 10
minutes. Their inclination appears more horizontal than vertical with
respect to the solar surface, while their orientation slightly changes
in some frames. Sometimes nearby mottles seem to merge forming a 
single structure or they split in two parts. Their shape and length clearly change from frame to frame as can be seen in Figure 1. 

The optical depth at the line center varies from 0 to 4 with a
mean value of 1.7 which is in agreement with the mean value found by
Lee et al. (2000). The center parts of the structures under
investigation show higher values, while their boundaries correspond to
lower ones. The source function varies between 0 and
0.15. For this parameter we found a mean value of 0.08, which is below
the values found by other authors \citep{tziotziou03,lee00}. The
highest values occur close to the footpoints of the mottles. The Doppler
width ranges between 0 and 0.88~\AA\, with a mean value of
0.52~\AA. Lee et al. (2000) found a mean value of 0.55~\AA\, for this
parameter. The velocity takes values between $-$25 and 25 km/s
indicating the presence of both downward and upward motion as seen in
some frames of our time series (see Figure 2). During our observation,
the footpoints of dark mottles mostly showed red-shift while their
upper parts showed both blue and red shift. The dominant motion is an upflow with a mean value of 1.64 km/s. 
\acknowledgements The Vacuum Tower Telescope is operated by the Kiepenheuer-Institut f\"ur Sonnenphysik, Freiburg, at the Spanish Observatorio del Teide of the Instituto de Astrof\'{\i}sica de Canarias.


\end{document}